# Observation of Enhanced Transformer Ratio in Collinear Wakefield Acceleration

C.Jing[1], A.Kanareykin[1], J.G.Power[2], M.Conde[2], Z.Yusof[2], P.Schoessow[1], and W.Gai[2]

[1.] Euclid Techlabs LLC, Solon, OH-44139

[2.] High Energy Physics Division, Argonne National Laboratory, Argonne, IL-60439

Abstract: One approach to future high energy particle accelerators is based on the wakefield principle: a leading high-charge *drive* bunch is used to excite fields in an accelerating structure or plasma that in turn accelerates a trailing low-charge *witness* bunch. The transformer ratio $R$ is defined as the ratio of the maximum energy gain of the witness bunch to the maximum energy loss of the drive bunch. In general, $R < 2$ for configurations in which the two beams traverse the accelerator along the same trajectory (collinear wakefield acceleration). A number of techniques have been proposed to overcome the transformer ratio limitation. We report here the first experimental study of the ramped bunch train (RBT) technique in which a dielectric loaded waveguide was used as the accelerating structure. A single drive bunch was replaced by two bunches with charge ratio of 1:2.5 and a separation of 10.5 wavelengths of the fundamental mode. An average measured transformer ratio enhancement by a factor of 1.31 over the single drive bunch case was obtained in this experiment.

In a wakefield accelerator, the fields generated by a leading, high-charge *drive* bunch (either a single drive bunch or a train of drive bunches) are used to accelerate a trailing, low-charge *witness* bunch. An important parameter that influences the performance of a wakefield accelerator is the transformer ratio $R \equiv$ (maximum energy gain of the witness bunch) / (maximum energy loss of the drive bunch) [1]. Therefore, to accelerate the witness beam to high energy it is desirable to make R as large as possible. For instance, consider a plasma wakefield accelerator in which a 10 GeV drive bunch decelerating at a rate of 10 GeV/m (due to its self-wake) will be spent in 1 m. Over the same distance, the witness bunch will be accelerated by R·10 GeV. (The beam loading by the witness beam is neglected since we assume its charge to be low.) *R* is less than 2 under very general conditions: linear media; a relativistic, longitudinally symmetric drive bunch; and identical paths through the system of both drive and witness beams [2, 3]. Some of the methods that can be employed to obtain $R > 2$ include: a triangular drive bunch longitudinal profile [4]; a train of Gaussian drive bunches of progressively increasing charge (Ramped Bunch Train, RBT) [5]; use of a proton drive beam so that the particles within the bunch can change positions during deceleration [6]; and nonlinear plasma dynamics [7].

In this Letter, we report on the observation of an enhanced transformer ratio in a collinear dielectric-loaded wakefield accelerator [8-10] by using the RBT technique. However, this technique is also applicable to other accelerators, most importantly to beam-driven plasma wakefield accelerators that can now be operated at extremely high gradients [11] but are still limited to $R < 2$ [12, 13]. In the RBT technique, the individual bunches in the train are symmetric, but the charge of each bunch and the interbunch spacing are determined by requiring that all bunches in the train experience the same maximum decelerating field. The RBT algorithm to achieve maximum transformer ratio enhancement has been fully discussed in refs. [14, 15], which emphasized that the two key parameters necessary to adjust in order to effect the transformer ratio enhancement are the optimized charge ratios and the spacing between bunches.

For a train of $N$ drive bunches, the requisite position $d_m$ of the $m^{th}$ bunch relative to the leading bunch is an odd multiple of $\lambda_0/2$, where $\lambda_0$ is the fundamental wavelength of the structure ($d_m = (m + 1/2) \lambda_0$, $m = 1, 2...N-1$). Each trailing bunch is located at an accelerating phase of the preceding bunch and its self-wakefield will cancel part of the accelerating field so the trailing bunch experiences a reduced deceleration. An optimized ramped-up charge ratio is then able to satisfy the requirement that all bunches in the train experience the same maximum decelerating field. The maximum accelerating field induced by the trailing bunch is partly canceled by the decelerating field of the preceding bunch but there is a net increase in the wakefield because the charge of each successive bunch is also increasing. As a result the transformer ratio increases with each successive bunch. Given the transformer ratio of the first bunch $R_1$ (determined by the ratio of the bunch length to the accelerating wavelength), the transformer ratio after the $n^{th}$ bunch is

$$R_n = R_{n-1} + (R_1 - 1)^{n-1} R_1 \qquad (1)$$

($n = 2, ... N$) assuming a ramped charge relation $Q_n = (1+R_{n-1})Q_1$. For example, if $R_1=2$, the transformer ratio increases as $R_n=nR_1$.

The experiment was carried out at the Argonne Wakefield Accelerator (AWA) facility located at Argonne National Laboratory (ANL) [16]. The AWA is based on an L-band photoinjector and booster linac; the bunch train consisting of two drive bunches and one witness bunch needed for the experiment are generated by splitting and delaying the UV laser pulse incident on the photocathode of the rf gun. The interbunch separations correspond to one rf period of the linac ($\lambda_L = 23 cm$) and are adjusted using optical delay lines. The accelerated bunches exit the linac and pass through a wakefield accelerator device and then into a magnetic dipole spectrometer.

A Dielectric-Loaded Accelerating (DLA) structure [8] was used to generate the wakefields for these measurements. The DLA is a dielectric tube enclosed by a conducting cylinder. The device built for this experiment has a total length of 40 cm, a vacuum channel radius of 5 mm and an outer radius of 6.34 mm. The permittivity of the dielectric is 16. The structure is described in detail in ref. [17]. The structure was instrumented with a field probe to detect the radial component of the wakefield $E_r$. The probe was located 82 mm downstream of the entrance of the DLA so that the direct wakefield signal could be separated from the reflection from the exit of the structure (an interval of a few nanoseconds).

The fundamental ($TM_{01}$) accelerating mode of this structure occurs at $f_0$= 13.625 GHz. The AWA linac operates at 1.3 GHz corresponding to an inter-bunch distance $d = \lambda_L = 10.5\lambda_0$ (where $\lambda_0$ = 22 mm at $f_0$), satisfying the requirement of half integer spacing between bunches for transformer ratio enhancement.

Because of the small rms bunch length of the drive beam available at the AWA linac (and its correspondingly broader frequency spectrum) the second higher order

(TM$_{02}$) accelerating mode at 39.25 GHz for this structure is also excited and contributes an additional 20% to the wakefield transformer ratio [3]. Leading and trailing drive bunches with charge 8 nC and 20 nC respectively were generated to satisfy the requirement of an equal decelerating field inside each bunch. Figure 1 shows a numerical simulation of the longitudinal wakefield based on the parameters of the experiment (including the attenuation caused by dielectric losses). Three modes (TM$_{01}$, TM$_{02}$, and TM$_{03}$) are considered in the simulation since higher frequency modes have very weak contributions to the generated wakefield (<1%) in this case. In the simulation, the longitudinal wakefield excited by the leading drive bunch (with measured rms bunch length of 1.5 mm) has a transformer ratio of 1.8, and 770 ps later (equivalent to 10.5 rf periods at $f_0$, or $\lambda_L$), the second bunch (rms bunch length of 2 mm) arrives at the maximum accelerating phase of the wake from the leading bunch. The larger charge of the second bunch depresses the resultant wakefield inside the bunch to a negative (decelerating) value. During the experiment, the charge ratio was adjusted so that both drive bunches experienced the same decelerating field, while simultaneously increasing the accelerating field beyond the single drive bunch value. A 1 nC witness bunch is launched at 1.54 ns ($2 \times \lambda_L$) after the first drive bunch to probe the accelerating wakefields generated either by the first drive bunch alone or both drive bunches. In Fig. 1, the simulated maximum decelerating fields inside both drive bunches are $W_{d1}^- = W_{d1+d2}^- = 1.6$ MV/m, while the maximum accelerating field after the first bunch is $W_{d1}^+ = 2.9$ MV/m and after the second bunch $W_{d1+d2}^+ = 4.9$ MV/m; an increase of a factor of 1.7. Thus, the transformer ratio increases correspondingly, from $R_1=1.8$ to $R_2=3$.

During the experiment, the longitudinal wakefield from the drive bunch is determined at a given phase by measuring the energy change of the witness bunch [10]. At the AWA facility, only one photoinjector was used to generate both drive bunches and witness bunch; all bunches have the same beam energy at the same launching phase. However, the high charge of the drive bunches (8-20 nC) and length of the DLA structure (0.4 m) result in energy changes of the witness bunch in the MeV range that can be measured in the spectrometer.

The bunch separation is adjusted optically by moving mirrors on translation stages in the delay lines. The optical delay/beam splitting system did not have the capability of adjusting the bunch delays continuously but rather in 2 mm increments. A key issue in this experiment is the accurate setting of the spacing between bunches to find the best approximation (a) to the $d1$-$d2$ delay yielding maximum R and (b) the witness beam delay resulting in maximum gradient. The initial spacing is roughly set up by keeping the same launching phase for the two drive bunches in the photoinjector. The final bunch spacing was obtained during the experiment by measuring the energy of the second bunch as a function of the delay; the maximum energy increase of the second bunch corresponds to the correct delay. Fig. 2 shows the measurement results of the normalized energy change of the second bunch due to the interaction with wakefields from the first drive bunch as a function of the spacing between them. The phase difference is converted to distance based on the $TM_{01}$ frequency $f_0$. The spacing for the maximum transformer ratio enhancement is indicated in Fig. 2, where the measured average energy of the first bunch is equal to the average energy of the accelerated second bunch ($W_{d1}=W_{d1+d2}$).

Figure 3 shows the observed probe signal and its frequency spectrum. Both the lowest accelerating mode (TM$_{01}$) at $f_0$ and the lowest dipole mode (HEM$_{11}$, driven by a small radial beam offset) at 12.4 GHz are detected; higher order modes are outside of the ~2 GHz bandwidth of the electronics. The measurement waveform clearly shows the signature of the launch of the second drive bunch. The slowly varying envelope of the wakefield signal results from the finite bandwidth of the detection system. The transformer ratio enhancement can be estimated from the ratio of the amplitudes $W_{d1+d2}^{+}/W_{d1}^{+} \approx 2$. The contamination by the HEM$_{11}$ mode (and the lack of a second orthogonally mounted probe in the DLA to completely diagnose it) makes it difficult to obtain a more accurate measurement using this technique.

The use of a witness beam to diagnose the fields provides a direct measurement of the transformer ratio enhancement. Fig. 4 shows the energy spectrum of the witness bunch under three different beam configurations. The mean energy (averaged over a number of measurements) of the witness bunch alone was 14.79±0.05 MeV; the bunch is accelerated to 15.40±0.05 or 15.59±0.04 MeV by the first or both drive bunches respectively. The witness bunch was launched at a delay of 1.54 ns with respect to the drive bunch, equivalent to 10.5×2× $\lambda_0$ (or 2×$\lambda_L$) bunch spacing. The delay between the witness and drive bunches was further adjusted by finding the highest energy shift of the witness bunch while varying the optical delay slightly around its nominal value. The average gradient at this delay was 1.67 MV/m, calculated by normalizing the 0.61 MeV shift of the witness beam caused by the first drive bunch to the 40 cm length of the DLA structure, and the group velocity of the wakefield (0.1$c$). The average gradient resulting from both drive bunches was similarly found to be 2.22 MV/m from the 0.80 MeV witness energy change. Since the decelerating fields inside each drive bunch are equal ($W^-_{d1}=W^-_{d1+d2}$), the wakefield transformer ratio enhancement factor can be calculated as

$$R_2/R_1 = \left(\frac{W^+_{d1+d2}}{W^-_{d1+d2}}\right) \bigg/ \left(\frac{W^+_{d1}}{W^-_{d1}}\right) = \frac{W^+_{d1+d2}}{W^+_{d1}} = 1.31 \pm 0.13. \qquad (2)$$

It is important to note that achieving an over-unity enhancement factor of the transformer ratio is a significant result and represents the first time that an experimental demonstration of this effect has been performed. In order to compare the experiment with the simulation results, it is necessary to take into account the finite length of the witness bunch in the experiment by convolving a gaussian of rms width 1.5 mm with the computed longitudinal wakefield, reordering the bins according to energy and finding the energy centroid of the distribution. The predicted finite bunch R enhancement at the witness phase is (2.95 MV/m)/(1.70 MV/m)=1.73. The difference between the predicted value of 1.73 and measured value of 1.31 is most likely caused by the 2 mm resolution of the optical delay adjustment; a single delay step off the peak corresponds to a decrease in the predicted transformer ratio enhancement factor by 0.43 units.

In conclusion, a transformer ratio enhancement in a dielectric collinear wakefield device was successfully demonstrated here for the first time through the use of a ramped bunch train. The measured transformer ratio enhancement is a factor of 1.31 larger than the single bunch case; based on comparison with numerical simulations the inferred transformer ratio in this experiment is 3. Future experiments are being planned to achieve R>>2 through the use of a four-bunch train.

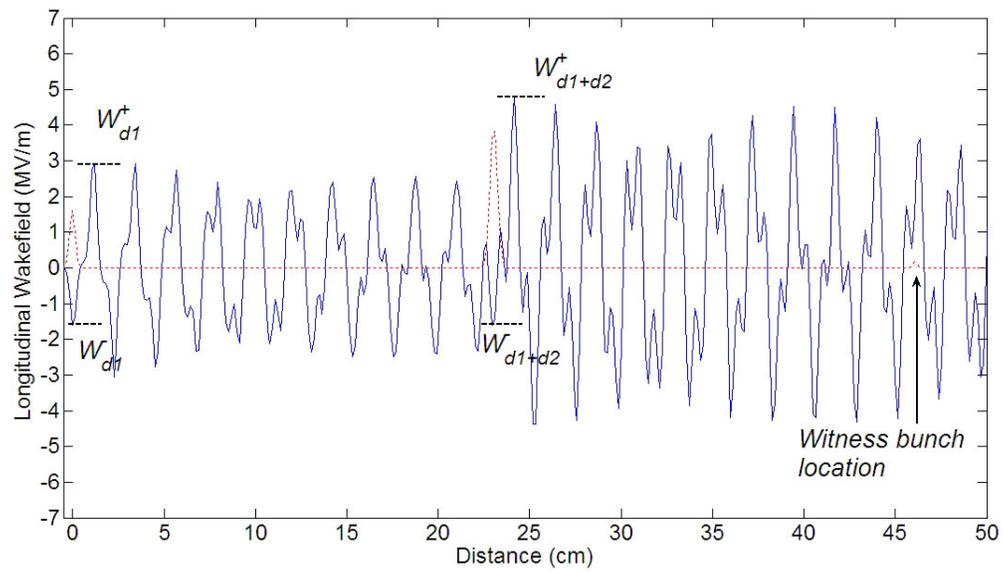

FIG. 1. Numerical simulation of the transformer ratio enhancement experiment. Dotted curve: beam current profile; Solid curve: longitudinal wakefield.

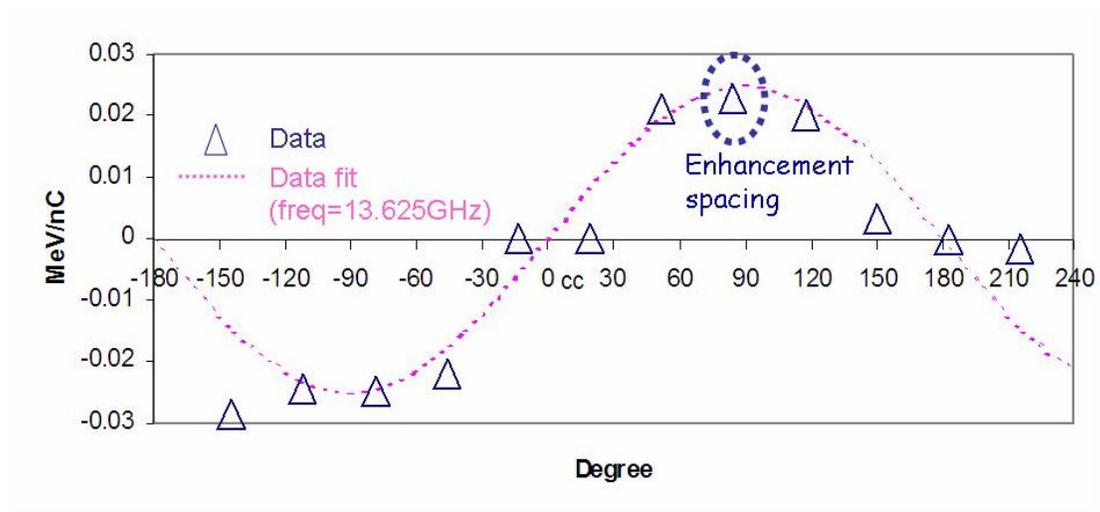

FIG. 2. Normalized energy change of the second drive bunch caused by the wakefield of the leading drive bunch as a function of their relative separation.

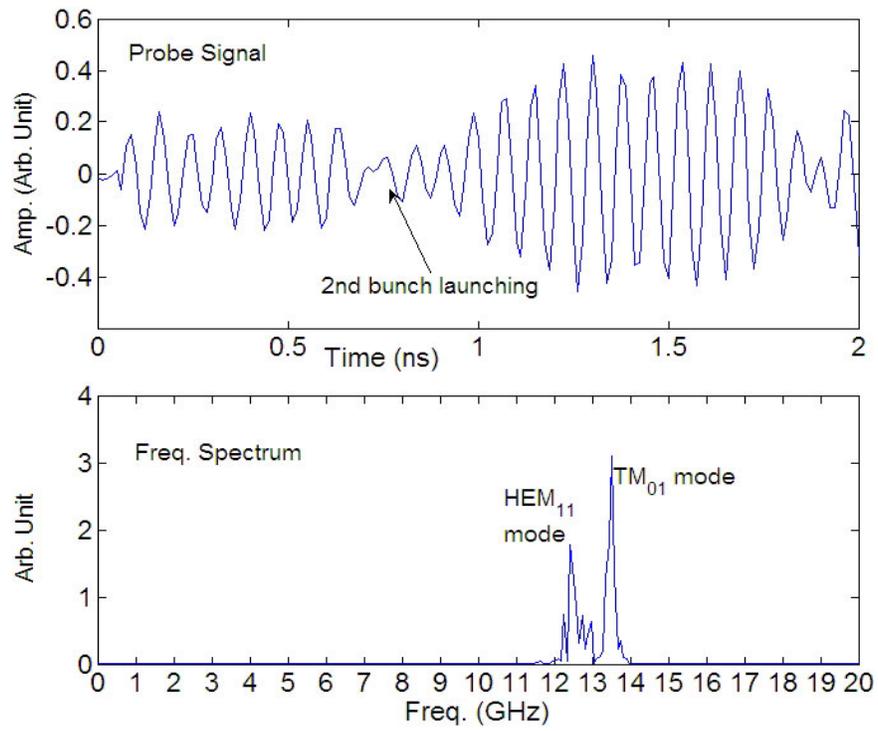

FIG. 3. The field probe signal from the RBT in the 13.625 GHz DLA structure and its frequency spectrum.

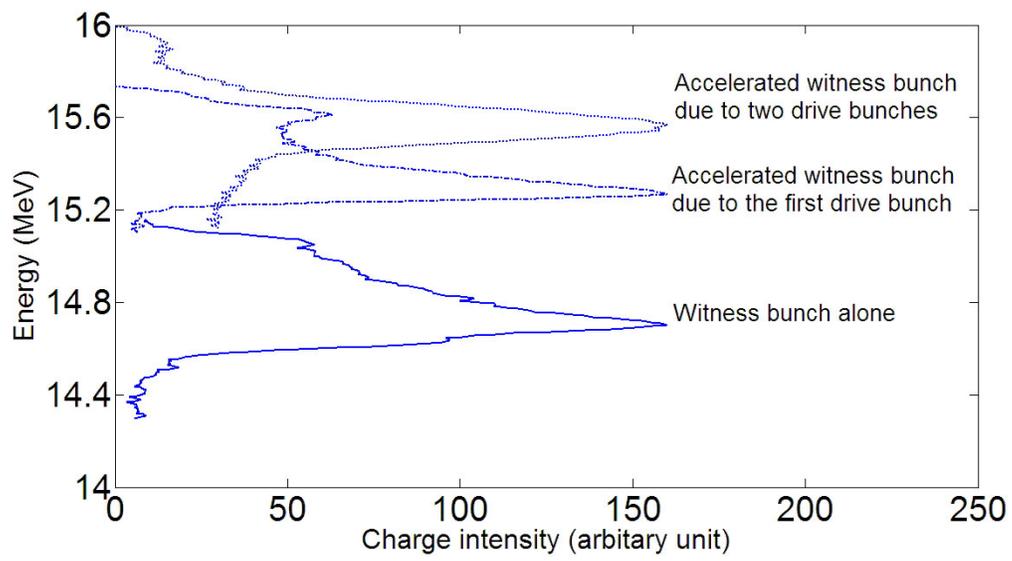

FIG. 4. Typical energy spectra of the witness bunch under three different drive beam configurations.